\documentclass[]{aa}  

\usepackage{graphicx}
\usepackage{txfonts}

\usepackage{siunitx}
\usepackage{tabularx}
\usepackage{multirow}
\usepackage{threeparttable}

\begin{document}

   \title{A systematic search for changing-look quasars in SDSS-II using difference spectra}

   \author{B. Potts
          \inst{1}
          \and
          C. Villforth\inst{1}\thanks{Email: c.villforth@bath.ac.uk}
          }

   \institute{Department of Physics, 
				 University of Bath, 
				 Claverton Down, 
				 Bath BA2 7AY, 
				 United Kingdom}
	
   \date{Accepted April 16, 2021}

 
  \abstract
    {
      `Changing-look quasars' (CLQs) are active galactic nuclei (AGN) showing extreme variability 
      that results in a transition from type 1 to type 2 AGN. The short timescales of these transitions 
      present a  challenge to the unified model of AGN and the physical processes causing these transitions remain poorly understood. CLQs also provide interesting samples for the study of AGN host galaxies since the central emission disappears almost entirely.
    }
    {
     Previous searches for CLQs have utilised photometric variability or SDSS classification changes to 
     systematically identify CLQs; this approach may miss lower luminosity CLQs. 
     In this paper, we aim to use spectroscopic data to asses if analysis difference spectra can be used 
     to detect further CLQs that have been missed by photometric searches.
    }
    {
      We searched SDSS-II DR 7 repeat spectra for sources that exhibit either a disappearance or appearance 
      of both broad line emission and accretion disc continuum emission by directly analysing the difference 
      spectrum between two epochs of observation.
    }
    {
      From a sample of 24,782 objects with difference spectra, our search yielded six CLQs within the redshift range 
      $0.1 \leqslant z \leq 0.3$, including four newly identified sources. 
      Spectral analysis indicates that changes in the accretion rate can explain the changing-look behaviour. 
      While a change in dust extinction fits the changes in the spectral shape, the timescales of the changes observed are too short for obscuration from torus clouds.
    }
    {
      Using difference spectra was shown to be an effective and sensitive way to detect CLQs. We recover CLQs an order of magnitude lower in luminosities than those found by photometric searches and achieve higher completeness than spectroscopic searches relying on pipeline classification. 
    }

   \keywords{Galaxies: active -- Accretion, accretion discs}

   \maketitle
%

\section{Introduction}

Active galactic nuclei (AGN) are known to be variable on timescales of months to years 
\citep[e.g.][]{MacLeod2010}. Their variability is generally associated with instabilities 
in the accretion disc. AGN are also found in two general types, type 1 and 2, the first 
of which shows strong broad emission lines and continuum emission, while the latter shows 
neither. These two types are generally explained as being due to orientation effects 
\citep{Antonucci93,UrryPadovani1995}. Only a small subset of AGN show continuum variability 
strong enough so that accretion disc emission appears or disappears entirely, accompanied with 
an appearance or disappearance of broad line emission. These objects are commonly referred 
to as `changing-look quasars' (CLQs). In this paper, we define changing-look behaviour to
be a clear transition between AGN types, irrespective of luminosity.

Changing-look quasars are relatively rare, but a notable well-researched 
example from the last 40 years is NGC 4151 
\citep[e.g.][]{Osterbrock77,Antonucci83,Penston84,Lyutyi84,Shapovalova10}, which has been 
observed `flickering`, as it transitioned back and forth between classification types. 
The first CLQ with quasar luminosity\footnote{
  We define the term quasar luminosity to mean bolometric luminosity $\geqslant 10^{44}$\,erg\,s$^{-1}$.
}, 
SDSS J015957.64+003310.5 (hereafter referred to as J0159+0033), was identified by \citet{LaMassa15}. 
Subsequently, a number of similar objects have been identified 
\citep[e.g.][]{Runnoe16,Ruan16,MacLeod16,Gezari17,Yang18,Ross18,Stern18, MacLeod19}.

There are a number of proposed scenarios that could explain changing-look behaviour. 
A popular explanation is that the observational variability is caused by a significant 
change in the accretion rate onto the central black hole. This drop in ionising flux causes a 
dimming of the broad line emission, while the narrow line region (NLR) remains unchanged due 
to the longer light travel time. Such a sudden change in ionising flux could be caused by 
instabilities in the disc or accretion state transitions \citep{NodaDone18}. Such a change 
could either be gradual, with AGN transitioning from high to low accretion states, or stochastic, 
with objects experiencing repeated extreme changes in accretion \citep[e.g.][]{Penston84,Elitzur14}. 
Alternatively, the movement of an isolated dust cloud in the torus can cause a change in extinction 
along the line of sight that would produce extreme broad line variability \citep{Nenkova08}. This 
mechanism is disfavoured for some cases \citep[e.g. J0159+0033,][]{LaMassa15} based on the timescale 
of the transition, but it could be responsible for some changing-look behaviour. Another alternative 
explanation for changing-look behaviour are tidal disruption events (TDEs) 
\citep[e.g.][]{Eracleous95,Merloni15}. \citet{Merloni15} propose that a luminous flare caused by 
the tidal disruption of a star near a supermassive black hole (SMBH) could have led to the observations 
of changing-look behaviour in J0159+0033. TDEs, however, cannot explain NLR emission on large scales 
ionised by the AGN. Studying changing-look behaviour can therefore give insights into accretion disc 
physics \citep{NodaDone18}, the structure of the obscuring torus \citep{Nenkova08}, as well as the 
incidence of tidal disruption events.

To date there have been four systematic searches for CLQs, yielding a total of 50 CLQs 
\citep{Ruan16,MacLeod16,Yang18,MacLeod19}. These data sets provide an invaluable resource for 
analyzing the likely physical mechanisms causing changing-look behaviour. Previous systematic CLQ 
searches have used either changes in the classification of objects in survey pipelines or on 
large photometric changes to identify objects of interest.

This focus on spectral classification changes and photometric variability favours the most obvious CLQs. 
In this work, we aim to assess the novel method of using difference spectra to identify CLQs, 
and to detect additional CLQs missed by previous searches of repeat spectra from SDSS-II.
We directly measure broad emission and continuum variability shown in the difference between two epochs,
and use them as sample selection criteria.

The outline of this paper is as follows. 
Sect.~\ref{SDSS_search} presents the method of using difference spectra to identify changing-look quasars. 
Sect.~\ref{CLQs} presents the results of the search, describing a sample of six CLQs, including four 
newly identified sources, the spectral decomposition method used to separate the host galaxy and the quasar 
components, and the emission line fitting procedure used to measure the broad emission lines and estimate 
the mass of the central supermassive black hole. 
Sect.~\ref{discuss} investigates variable accretion rate and variable obscuration as explanations for the six CLQs. 
Sect.~\ref{conc} summarises and makes concluding comments. 
Where needed, we adopt a flat $\Lambda$ cold dark matter cosmology, 
where H$_0$ = 69.6 km\,s$^{-1}$ Mpc$^{-1}$ and $\Omega_{\text{M}}$ = 0.286.


\section{SDSS spectroscopic data search}
\label{SDSS_search}

Our search for CLQs directly compared repeat spectroscopy from SDSS DR7 by constructing a `difference spectrum' 
for each object in our sample to identify changing-look behaviour. By analysing the difference spectrum in our 
search methodology, we hoped to isolate and measure the variable component between epochs of observation. 
We aimed to show that our search is sensitive to changing-look behaviour by identifying new CLQs from SDSS-II.

\subsection{Spectroscopic data}

We used the spectroscopic data from the final SDSS-I/II data release 7 (DR7). DR7 contains 1,630,960 spectra, including 
929,555 galaxy spectra and 121,363 quasar spectra \citep{Abazajian09}. SDSS-I/II spectra are captured using a 3" 
diameter fibre and have a wavelength coverage of \numrange{3800}{9200}\,\AA\:at a resolution of $R\approx 2000$ 
\citep{York00}. We used the classification made by the SDSS spectroscopic pipeline \citep[see][]{Stoughton02} to 
select objects classified as either a galaxy (`GALAXY') or a quasar (`QSO'). We did not rely on changes in 
classification between repeat spectra. We included all objects with at least two different epochs of observation.

We chose SDSS DR7 to ensure all spectra were captured with only the SDSS Legacy spectrograph. 
This removes any potential false positive detections due to instrumental changes. 
Since the aim of this work was to show how spectroscopic searches can be used to expand on photometric searches, 
rather than create large samples of CLQs, SDSS DR7 data alone is used.
All observed spectra shown were corrected for galactic extinction using the extinction law of \citet{Fitzpatrick99}, 
and the maps of \citet{Schlegel98}.

\newcolumntype{Y}{>{\centering\arraybackslash}X}

\begin{table}
	\centering
   \caption{
     Changing-look identifier thresholds. 
   }
	\label{tab:thresholds}
    \begin{tabularx}{\columnwidth}{ lc *{2}{Y} }
		\hline
		\multirow{3}{*}{Detection strength} & Continuum & \multicolumn{2}{c}{Peak emission line flux} \\
		& ($10^{-17}$\,erg\,s$^{-1}$ & \multicolumn{2}{c}{($10^{-17}$\,erg\,s$^{-1}$\,cm$^{-2}$)} \\
        & cm$^{-2}$\AA$^{-2}$) & H$\alpha$ & H$\beta$ \\
		\hline
        
		Strong & >\,\num{-2e-2} & >\,\num{3.5} & >\,\num{5.25} \\
        Intermediate & >\,\num{-1.8e-3} & >\,\num{1.8} & >\,\num{2.7} \\
        Weak & >\,\num{-6e-4} & >\,\num{0.8} & >\,\num{1.2} \\

		\hline
	\end{tabularx}
  \tablefoot{
    Continuum refers to the mean slope of the continuum in the rest wavelength range \numrange{3500}{3700}\,\AA. 
     Objects were selected by our algorithm for manual review if they displayed at least one strong CLQ feature and one weak CLQ feature. 
     Objects that displayed only one CLQ feature are discarded. 
     Objects with two intermediate strength features were also selected for manual review.
  }
\end{table}

\subsection{Sample selection}
\label{sec:select}

A super-set of candidate CLQs was selected using the following criteria. 
Firstly, each object must have repeat spectra in SDSS DR7 in at least two different epochs.
Secondly, each object must be classified as either GALAXY or QSO in all epochs of observation, rejecting stars.
Finally, each object must have redshift, $0.1 \leq z \leq 0.4$ to ensure that both H$\alpha$ and the blue 
continuum at $\sim$ 3500\ \AA\, are visible.
The resulting super-set consisted of 55,116 spectra from 24,782 objects.

Changing-look behaviour is defined as a clear transition between AGN types. 
An AGN exhibiting a convincing type transition in the optical possesses at least two features: 
a strongly brightening or dimming continuum and broad line variability across its multi-epoch spectra. 
We identified CLQs by quantifying these two changing-look features using the difference spectrum, 
$\left|\Delta f_{\lambda}\right|=\left|f_{\text{epoch 2}}-f_{\text{epoch 1}}\right|$. 
The observed spectra were smoothed using a Gaussian filter before computing the difference spectrum 
to reduce the effect of noise. 
We modelled the continuum as a third-degree polynomial and fitted to the entire wavelength range available 
on the difference spectrum. 
This fit was used to subtract the continuum and isolate emission lines.

As an initial step, we preselected objects showing considerable continuum variability at 
\numrange{4100}{5500}\,\AA\, (corresponding to the SDSS $g$-band). 
We selected objects with a variability exceeding $3 \times 10^{-18}$\,erg\,s$^{-1}$\,cm$^{-2}$\AA$^{-1}$ 
corresponding to $\Delta mag \sim 0.16$ at a magnitude of 21.
If variability is caused by changes in the accretion rate, we expect the difference spectrum continuum of 
a CLQ to resemble a Shakura-Sunyaev accretion disc power law \citep{Shakura73}.

\begin{table}
	\centering
	\caption{
      CLQ sample selection cuts.
   }
	\label{tab:selection}
   \begin{tabularx}{\columnwidth}{l @{\extracolsep{\fill}} cc}
		\hline
		Selection & Total & Changed class\\
		\hline
		SDSS objects & 1,053,144 & ...\\
        GALAXY or QSO & 954,447 & ...\\
        $0.1<z<0.4$ & 439,110 & ...\\
		Repeatedly observed objects & 24,782 & 184\\
		Selected by algorithm & 941 & 38 \\
        CLQs (new) & 6 (4) & 3 (1)\\
		\hline
	\end{tabularx}
  \tablefoot{
    The sample was reduced to a manually manageable size with a series of cuts from top to bottom. 
    The column marked `Changed class' indicates objects that changed object classification between 
    GALAXY and QSO, or vice verse, in the same way as the sample selection by \citet{Ruan16}.
  }
\end{table}

In a variable obscuration scenario, blue light would be attenuated more than red 
\citep{VandenBerk04, Wilhite05, Schmidt12}. 
Similarly, in a variable accretion rate scenario stronger changes are expected at short wavelength due 
to the spectrum of the accretion disc. 
Consequently, we searched difference spectra for a strong blue continuum.

We used the mean slope of the polynomial-continuum in the wavelength range \numrange{3500}{3700}\,\AA\, 
to characterise the strength of the continuum variability. 
Three threshold values, listed in Table \ref{tab:thresholds}, quantitatively define weak, intermediate, 
and strong detections of continuum variability.

Changing-look broad line variability is measurable as it manifests itself in the difference spectrum as 
apparent broad emission lines. 
We specifically searched for the strongest optical broad emission lines: H$\beta$; and H$\alpha$, by 
extracting our modelled continuum from the difference spectrum, and then identifying the highest 
monochromatic flux in the rest wavelength ranges \numrange{4856}{4866}\,\AA\: and 
\numrange{6558}{6568}\,\AA, respectively. 
For a CLQ, we expect only broad lines to appear in the difference spectrum since narrow lines are not 
expected to vary over short timescales due to extra light travel time from the continuum source to the 
narrow line region, relative to the broad line region. 
So, at this stage we were not concerned with measuring the width of the emission in the above wavelength 
range, only the flux. 
Three threshold values each for the detected H$\alpha$ and H$\beta$ emission line flux, listed in 
Table \ref{tab:thresholds}, were used to define weak-, intermediate-, and strong-cases of broad emission 
line variability\footnote{
  All threshold values were optimised before searching the entire super-set using a subset of 
  184 objects whose SDSS pipeline classification changed between GALAXY and QSO across at least 
  two epochs. Thresholds were chosen so that the algorithm selected 38 objects of interest for visual 
  inspection from said subset.
}.
Objects that displayed both changing-look indicators (except those with only weak detections of both continuum 
and broad line variability), or a strong-case of one changing-look indicator, were selected as changing-look 
candidates. 

\begin{table*}
	\centering
	\caption{
    CLQs found.
  }   
  \begin{threeparttable}
	\label{tab:changinglook}
  \begin{tabularx}{\textwidth}{lc *{3}{Y} *{2}{c} Y}
		\hline
		\multirow{2}{*}{Name (SDSS J)} & \multirow{2}{*}{Source} & \multirow{2}{*}{MJD$_{\text{epoch 1}}$} & \multirow{2}{*}{MJD$_{\text{epoch 2}}$} & \multirow{2}{*}{z} & log$_{10} (\lambda$L$_{5100})$ & \multirow{2}{*}{$\left|\Delta f_{\lambda}\right| \propto \lambda^{-\beta}$} & \multirow{2}{*}{Type}\\
		 &  & & & & (erg\,s$^{-1}$) &  & \\
		\hline
		082323.89\,$+$\,422048.3 & This paper & 52266 & 54524 & 0.152 & 43.40$\pm$0.001 & 2.164$\pm$0.030 & off\\
        172322.31\,$+$\,550413.8 & This paper & 51813 & 51997 & 0.295 & 44.08$\pm$0.001 & 2.647$\pm$0.028 & on \\
		012648.08\,$-$\,083948.0 & 1 & 52163 & 54465 & 0.198 & 43.92$\pm$0.001 & 1.727$\pm$0.023 & off \\
		082942.67\,$+$\,415436.9 & This paper & 52266 & 54524 & 0.126 & 44.06$\pm$0.001 & 1.631$\pm$0.019 & on \\
        000236.25\,$-$\,002724.8 & This paper & 51791 & 52559\tnote{a} & 0.291 & 43.68$\pm$0.126 & 1.993$\pm$0.042 & off \& on \\
        135855.83\,$+$\,493414.2 & 2 & 53438 & 54553 & 0.116 & 43.24$\pm$0.001 & 1.330$\pm$0.046 & on \\
		\hline
	\end{tabularx}
  \tablefoot{
    log$_{10} (\lambda$L$_{5100})$ is the bright-state spectrum continuum luminosity, 
    $\left|\Delta f_{\lambda}\right| \propto \lambda^{-\beta}$ is the best-fitting power law index to the 
    difference spectrum, and Type is the type of changing-look behaviour, with off indicating disappearing 
    broad emission lines, and on indicating appearing broad emission lines.\\
    \tablefoottext{a}{
      The object J000236.25-002724.8 was re-observed by BOSS giving a third epoch of observation on MJD: 55477. 
    }
  }
  \tablebib{
    (1) \citet{Ruan16}; (2) \citet{Yang18}.
  }
  \end{threeparttable}
\end{table*}

In summary, our selection algorithm first separated the continuum for emission lines, then accepted objects 
with continuum variability in the range \numrange{4100}{5500}\,\AA\, greater than $\Delta mag \sim 0.16$ at 
a magnitude of 21. 
We then quantified the changing-look indicators, blue continuum variability and broad emission line variability, 
and applied thresholds listed in Table \ref{tab:thresholds} to complete our search.

The search yielded 941 CLQ candidates. 
We visually inspected all candidate spectra, rejecting quasars that show variability, but do not exhibit an AGN 
type transition.
We identified a total of six CLQs, four of which are newly identified and evidence the sensitivity of the
systematic difference spectrum search methodology.
Two CLQs, J135855.83+493414.2 (hereafter referred to as J1358+4934), found by \citet{Yang18}, and 
J012648.08-083948.0 (hereafter referred to as J0126-0839), found by \citet{Ruan16}, were recovered. 
\citet{Yang18} finds two other CLQs from SDSS Legacy repeat spectra \citep[see Table 5 of][]{Yang18}, 
and these were also recovered by the sample selection algorithm, yet excluded from the analysis in the 
interest of only including those objects displaying the clearest and strongest changing-look behaviour. 
Sample selection cuts that were made are shown in Table~\ref{tab:selection}.

To assess the validity of the sample selection methodology, we also applied the sample selection algorithm 
to the SDSS Legacy and BOSS spectra of the \citet{MacLeod16} CLQs. 
Our sample selection algorithm recovered seven of the ten objects identified by \citet{MacLeod16}. 
The three missing objects were not recovered using our method as they were either outside our 
redshift range criterion or missing data in the relevant spectral range.


\section{The changing-look quasars}
\label{CLQs}

Our difference spectrum search method identified six CLQs in SDSS-II, including four newly found sources using the difference spectrum method developed here. 
Two CLQs `turned off' (disappearing broad lines and continuum), three `turned on' (appearing broad 
lines and continuum), and one was observed to `turn off and turn on' (disappearing and reappearing 
broad lines and continuum, after an additional BOSS spectrum was identified). 
The properties of the final sample of CLQs are listed in Table \ref{tab:changinglook}. 
The objects have redshifts $0.116 \leq z \leq 0.295$.

\subsection{Turning off changing-look quasars}

\begin{figure*}
	\includegraphics[width=\textwidth]{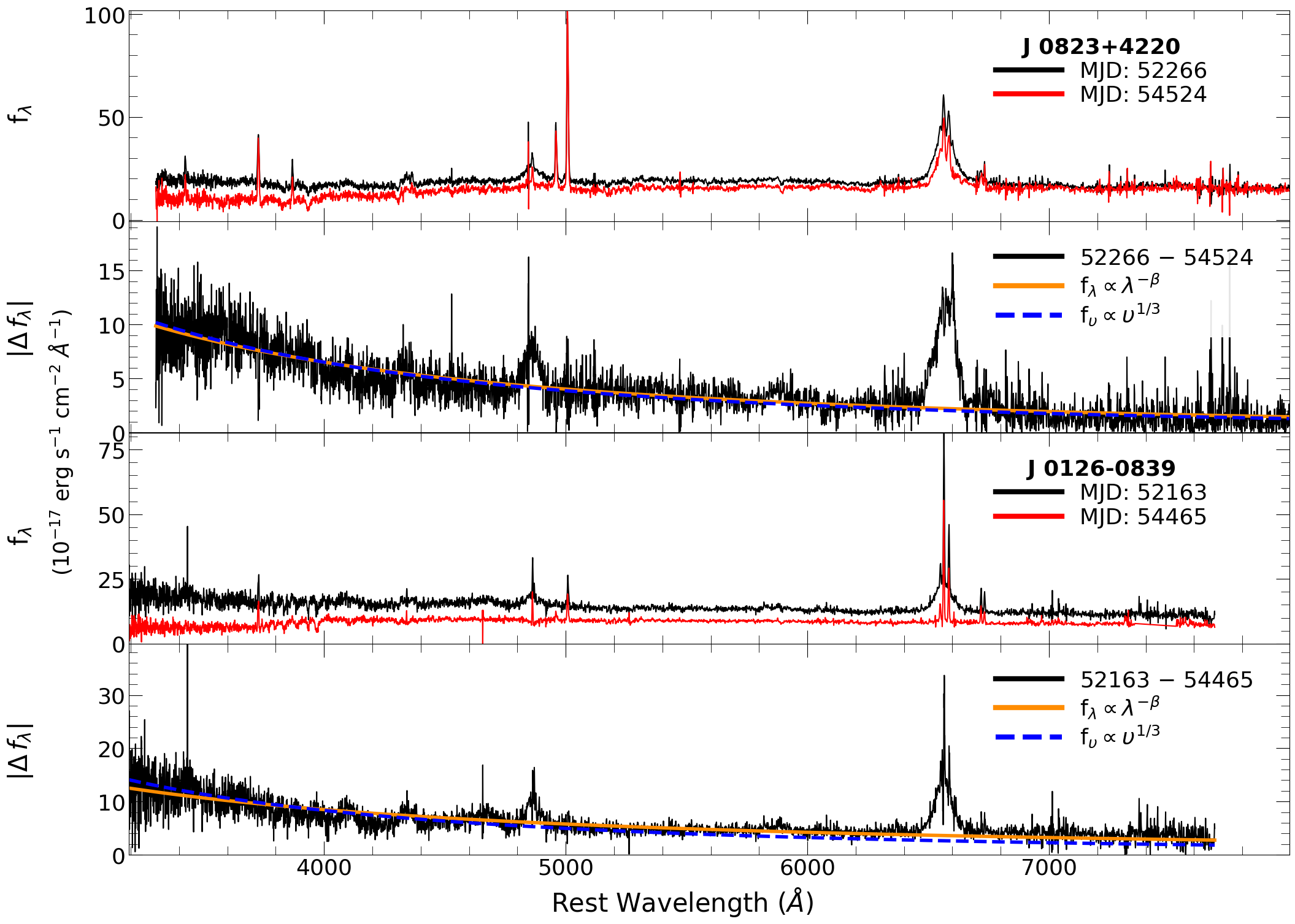}
    \caption{
      Observed spectra and the difference spectrum of the turning off CLQs. 
      The quasars shown are J0823+4220 (top two panels), and J0126-0839 (bottom two panels). 
      Upper: the observed optical spectra, with the first epoch (black) and second epoch (red) both shown. 
      Lower: the difference spectrum  $\left|\Delta f_{\lambda}\right|$ (black) is shown, demonstrating how 
      the CLQs were identified, and the resemblance of the variable component to an accretion disc. 
      A fitted Shakura-Sunyaev accretion disc power law $f_{\upsilon} \propto \upsilon^{1/3}$ (dashed-blue) 
      and a best-fit power law to the continuum (orange) is shown.
    }
    \label{fig:Off}
\end{figure*}

The two CLQs that turned off are displayed in Fig. \ref{fig:Off}. 
In the first epoch of observation of both objects, the AGN classification is type 1, and the H$\beta$ 
broad line completely disappears by the second epoch of observation.

The object J082323.89+422048.3 (hereafter referred to as J0823+4220) is shown in the top panel of Fig. \ref{fig:Off}. 
In its second epoch of observation, a relatively strong H$\alpha$ broad line remains, which characterises 
its AGN type as type 1.9. The transition happened on a timescale of 2258 days (1950 days rest frame).

The object J0126-0839, which was first found by \citet{Ruan16}, is shown in the bottom panel. 
J0126-0839 has the most extreme changing-look behaviour observed in the sample as both H$\beta$ and 
H$\alpha$ completely disappear. 
A disappearing H$\gamma$ broad line is also evident, particularly in the difference spectrum (black curve). 
The AGN has transitioned to type 2 in its second epoch of observation on a time scale of 2302 days (1902 
days rest frame).

\subsection{Turning on changing-look quasars}

The three CLQs that turned on are displayed in Fig. \ref{fig:On}. 
In the second epoch of observation of all three objects, the AGN classification is type 1, as strong 
H$\beta$ and H$\alpha$ broad emission lines have emerged. 
We note that all three displayed asymmetry in their bright-state broad emission of both H$\beta$ and 
H$\alpha$, the broad emission lines show excess emission in the red wing. 
Similar asymmetry is not exhibited in the CLQs that turned off.

The object J172322.31+550413.8 (hereafter referred to as J1723+5504), shown in the top panel of Fig. 
\ref{fig:On}, transitioned from a type 1.8\footnote{
  type 1.5 objects are defined as intermediate between type 1 and type 2 with easily apparent broad H$\beta$ emission. 
  type 1.8 is intermediate between type 1.5 and type 2, with a weaker H$\beta$ than type 1.5 yet still 
  visible \citep{Osterbrock77}.
} to type 1 AGN in just 184 days (142 days rest frame). 
To our knowledge, this is the fastest observed changing-look behaviour in a distant AGN with quasar luminosity. 
In its first epoch the presence of a weak broad H$\beta$ characterises J1723+5504 as an intermediate type 1.8. 
The H$\beta$ and H$\alpha$ broad lines clearly emerge by the second epoch. 
The broad H$\beta$ and H$\alpha$ are asymmetric in the bright-state epoch, with an extended red wing.

The object J082942.67+415436.9 (hereafter referred to as J0829+4154) is shown in the middle panel. 
Similar to J1723+5504, it is in an intermediate state in its first epoch: type 1.8. 
The observed broad components of H$\alpha$ and H$\beta$ appear to be either asymmetric with a red wing, 
as with that of J1723+5504, or are substantially redshifted with respect to their narrow components, 
in both the dim- and bright-state. 
The transition happened on a timescale of 2258 days (2005 days in the rest frame).

The object J1358+4934, which was first identified by \citet{Yang18}, is shown in the bottom panel. 
J1358+4934 shows the most extreme turning on changing-look behaviour in the sample as it transitions 
from AGN type 2 to type 1. 
To a less obvious extent than J1723+5504 and J0829+4154, J1358+4934 is also asymmetric in the emerging 
H$\alpha$ and H$\beta$ emission with a red wing in its bright-state spectrum. 
The transition happened on a timescale of 1115 days (999 days in the rest frame).

\subsection{Turning off and on changing-look quasar}

The multi-epoch spectra and difference spectra of J000236.25-002724.8 (hereafter referred to as J0002-0027) 
are shown in Fig. \ref{fig:Off_and_on}. 
The object was observed twice in SDSS-II, shown by the black and red curves in the top panel of 
Fig. \ref{fig:Off_and_on}, and was identified from these spectra. 
The continuum dimmed, the H$\beta$ broad line disappeared, and the H$\alpha$ broad line diminished, 
yet remained strong relative to H$\beta$, as the object turned off in transitioning from type 1 to 
type 1.9. The object exhibited this changing-look behaviour in 768 days (587 days in the rest frame). 
J0002-0027 was re-observed by SDSS-III BOSS, giving us an additional third epoch of observation (blue curve). 
The re-appearing broad emission lines and continuum from the third epoch almost perfectly match that of the 
first epoch. 
The only major observational difference to the first epoch spectrum is a blue excess in the broad H$\alpha$ 
emission in the BOSS spectrum. 
No such excess is seen in H$\beta$ and the width and shape of the line is otherwise unchanged, as shown by 
the bottom-right panel of Fig. \ref{fig:Off_and_on}. 
Overall, the object was observed turning off and on in 3686 days total (2818 days in the rest frame).

\subsection{Spectral analysis}

In the following section, we describe the spectral analysis performed to characterise the shape of the 
variable component, fit narrow and broad emission lines and measure black hole masses and Eddington ratios.

\subsubsection{Continuum variability}

\begin{figure*}
	\includegraphics[width=\textwidth]{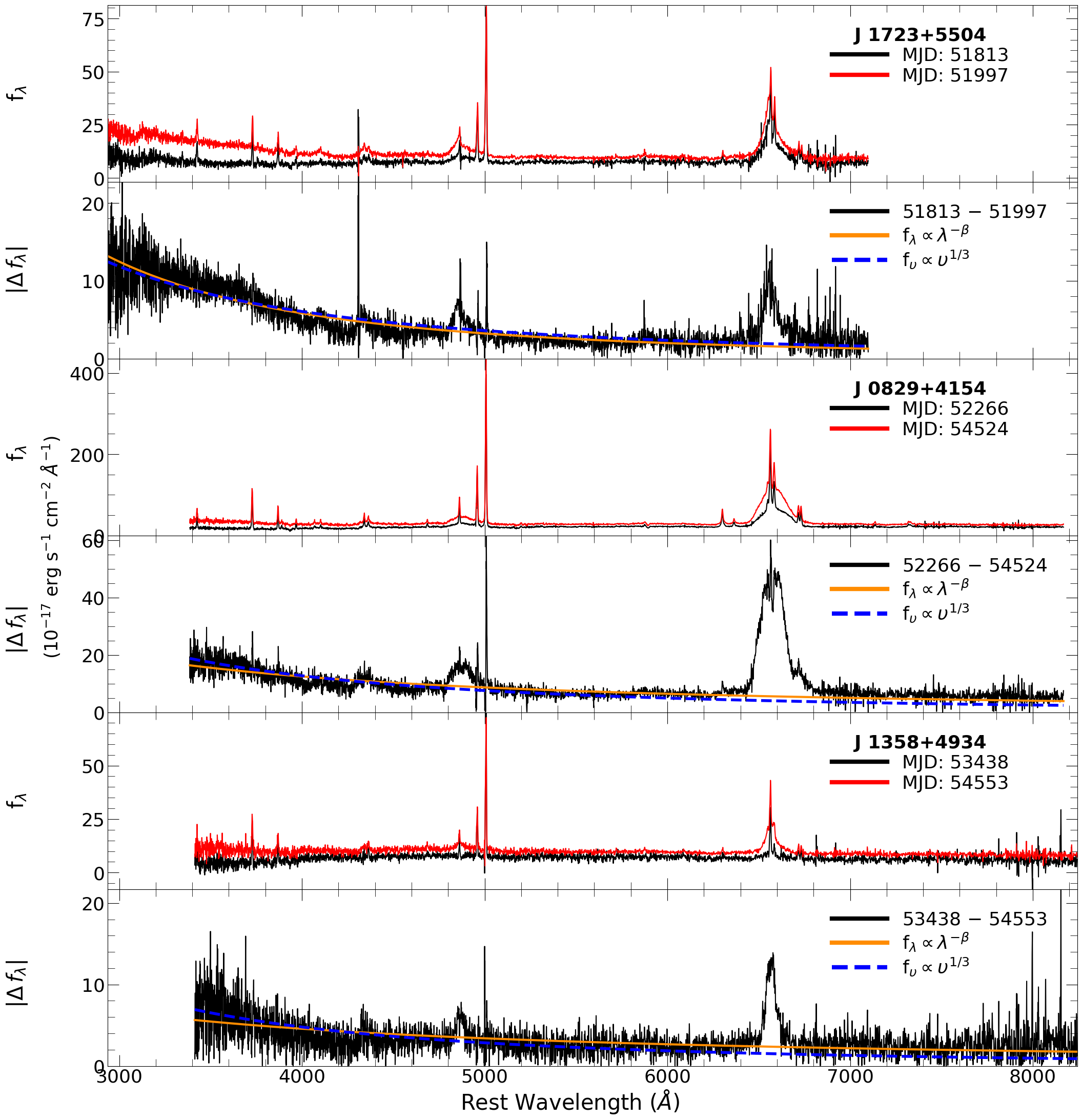}
    \caption{
      Observed spectra and the difference spectrum of the turning on CLQs. 
      The turning on CLQs are shown in the same format as Fig.~\ref{fig:Off}. 
      The quasars shown are J1723+5504 (top two panels), J0829+4154 (middle two panels), and J1358+4934 (bottom two panels).
    }
  \label{fig:On}
\end{figure*}

\begin{figure*}
	\includegraphics[width=\textwidth]{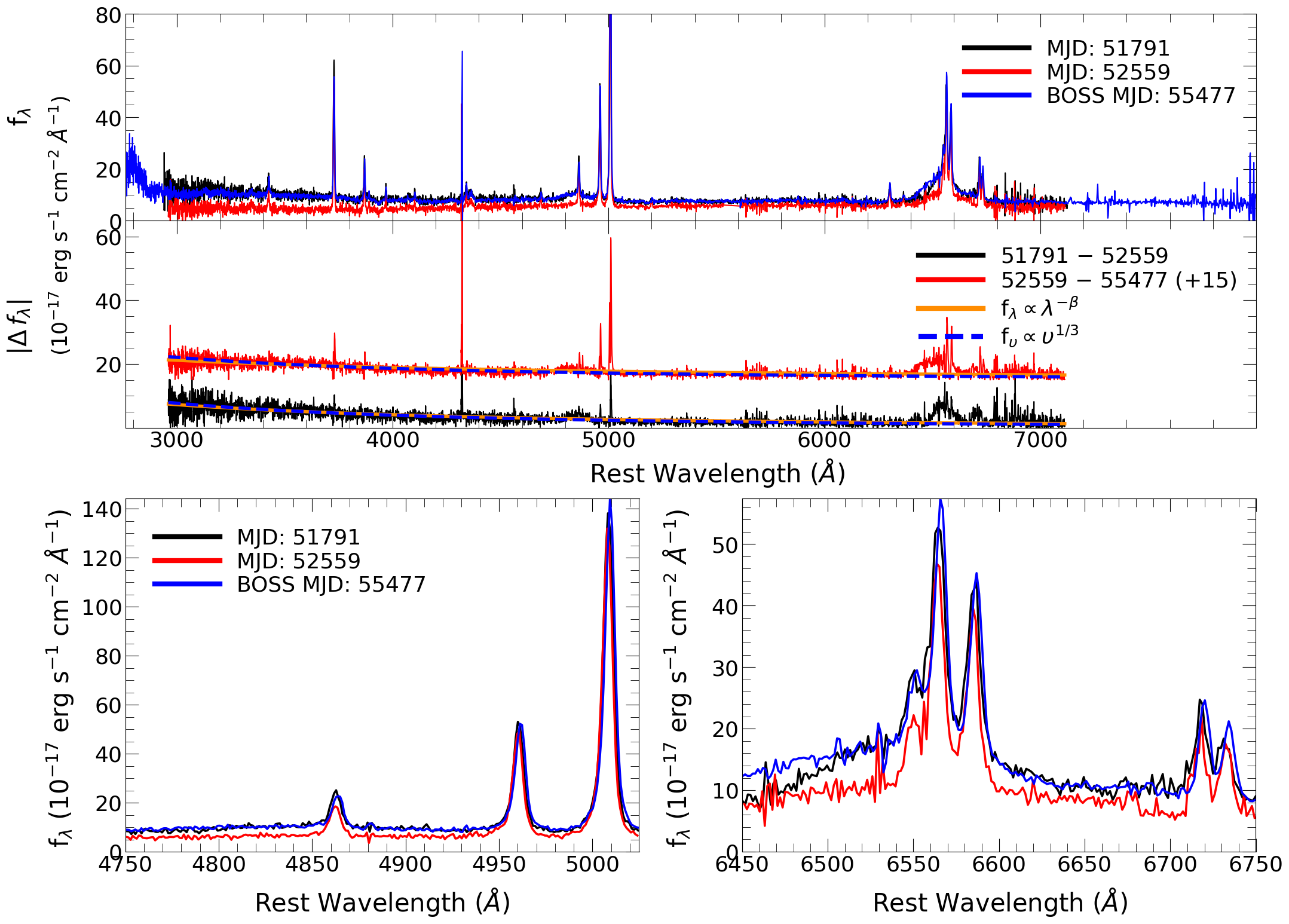}
    \caption{
      Observed spectra and the difference spectra of the turning off and on CLQ J0002-0027. 
      Top: the three observed spectra, including the extra BOSS spectrum (blue curve), are 
      shown in the SDSS Legacy rest wavelength range of the object. 
      The epochs for the black, red, and blue spectra are MJD\,=\,51791, 52559, and 55477, respectively. 
      Middle: The difference spectra of the first and second epoch (black curve), and the second and 
      third epoch (red) are shown on the same flux scale, with a constant added to the latter. 
      Bottom: A zoomed in view of the observed spectra in the H$\beta$ (left) and H$\alpha$ (right) regions is shown.
    }
  \label{fig:Off_and_on}
\end{figure*}

The difference spectrum (black curve in alternating panels of Figs. \ref{fig:Off} and \ref{fig:On}, 
black and red curves in the middle panel of Fig. \ref{fig:Off_and_on}) shows the variable component, 
exhibiting a blue continuum. 
We fitted the spectrum with a power-law $\left|\Delta f_{\lambda}\right| \propto \lambda^{-\beta}$ (orange curve) 
as well as a standard thin accretion disc power-law of form $f_{\upsilon} \propto \upsilon^{1/3}$ (dashed blue curve) 
\citep{Shakura73}. 
For the fit, we masked regions around prominent Balmer emission lines 
(H$\alpha$: \numrange{6420}{6800}\,\AA; H$\beta$: \numrange{4770}{5050}\,\AA; H$\gamma$: \numrange{4280}{4400}\,\AA; 
and H$\delta$: \numrange{4050}{4140}\,\AA for J0126-0839). 
The difference spectrum of all six CLQs was well fitted by a standard thin disc model. 
The variability is therefore consistent with changes in the accretion rate.

\subsubsection{Spectral decomposition}
\label{sec:decomp}

We intended to derive the black hole mass and characterise the source of ionisation in the narrow line region. 
In order to isolate the AGN continuum, and the broad and narrow emission lines, we decomposed the observed 
spectra of the CLQ sample into host-galaxy and quasar components. 
Additionally, we used the decomposed spectra to determine if variable obscuration of the disc component 
can explain the observed behaviour.

To decompose the spectra, we used a modified version of the method by \citet{VandenBerk06}, who fitted spectra 
with quasar and galaxy eigenspectra. 
Specifically, we used the first five SDSS galaxy and the first five SDSS quasar eigenspectra described and 
made available by \citet{Yip04} and \citet{Yip04b}, respectively. 
We fitted linear combinations of the eigenspectra, with their coefficients as ten free parameters, to the 
observed spectra using a $\chi^2$ minimisation. 
\citet{Yip04,Yip04b} showed that their galaxy and quasar eigenspectra gave a statistical coverage of 
98.37\% and 99.81\% of their respective SDSS samples used to generate them. 
We masked the H$\beta$ and H$\alpha$ regions (\numrange{4780}{5080}\,\AA\;and \numrange{6270}{6800}\,\AA) 
due to the complexity of the broad emission lines. 
Aside from broad line masking, the decomposition was performed for the maximum available rest wavelength 
range of each object. 
All SDSS-II spectra are observed using a 3" diameter fibre, and we did not expect any intrinsic variability 
in the host galaxy over the relatively short timescales of our sample, meaning the observed host component 
should remain constant across repeat spectra. 
Hence, we decomposed repeat SDSS-II spectra of each object simultaneously while constraining the galaxy 
eigenspectrum coefficients as equal across all epochs of observation.

\begin{figure*}
	\includegraphics[width=\textwidth]{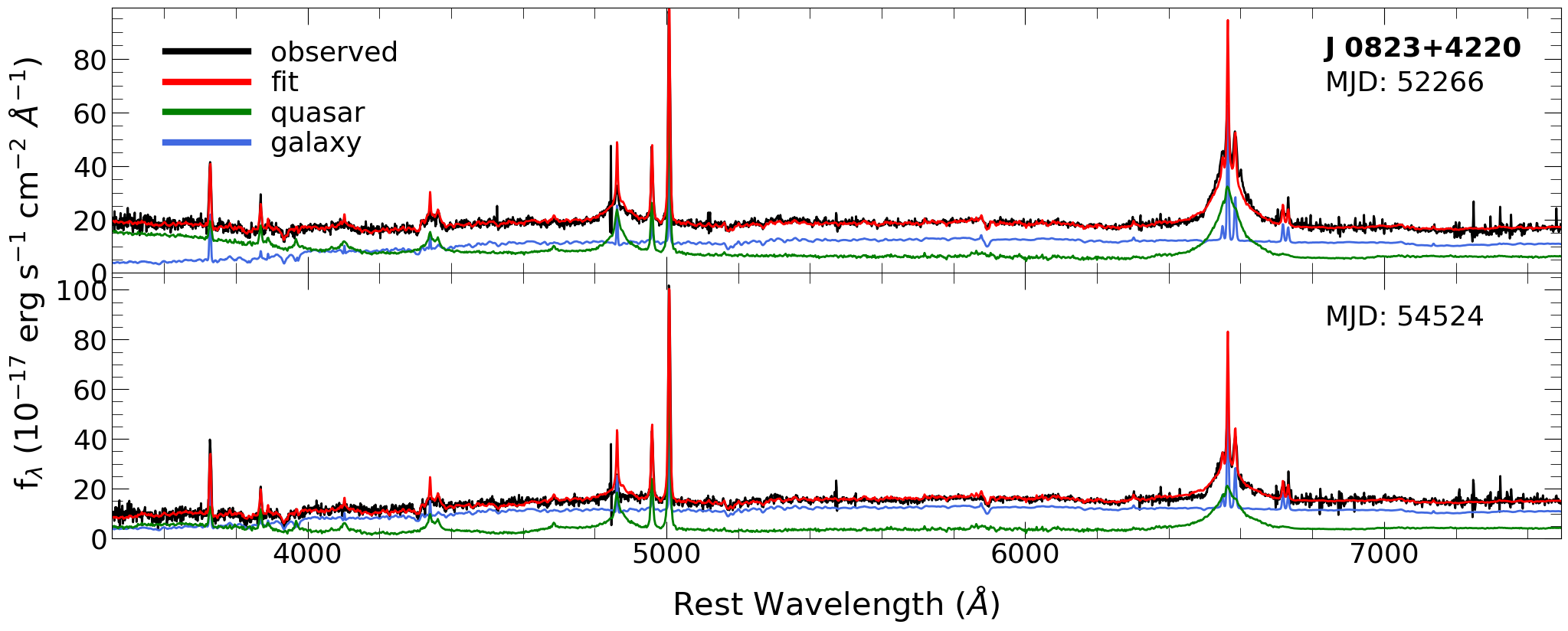}
    \caption{
      Spectral decomposition of the CLQ J0823+4220, which turned off. 
      The fitted galaxy (blue curve) and quasar (green) components, and their sum (red) from the best-fit 
      eigenspectrum coefficients (found as per Sect. \ref{sec:decomp}) are shown. 
      Top: the earlier epoch (MJD = 52266) bright-state spectrum. Bottom: the later epoch (MJD = 54524) 
      dim-state spectrum.
    }
    \label{fig:spec_decomp}
\end{figure*}

The result of decomposition for each spectrum were isolated galaxy (blue curve) and quasar (green) components, 
shown in Fig. \ref{fig:spec_decomp} for J0823+4220 with their sum (red) as an example result.
Similarly to Fig. \ref{fig:spec_decomp}, the other spectra were fitted well by our model. 
J0126-0839, which transitioned from a type 1 to type 2 AGN (the most extreme changing-look behaviour 
in the sample), was best fitted with zero quasar component in its second epoch spectrum.

\subsubsection{Broad \& narrow emission line fitting}
\label{sec:lines}

To measure the broad and narrow emission lines of only the quasar component of each spectrum, we used 
the host-galaxy component (blue curve in Fig. \ref{fig:spec_decomp}) yielded by the spectral 
decomposition of Sect. \ref{sec:decomp} and subtracted it directly from each observed spectrum 
(black curve in Fig. \ref{fig:spec_decomp}).
We chose to subtract the host galaxy from the observed to preserve as much detail from the observed 
spectrum as possible, rather than simply using the quasar component from Sect. \ref{sec:decomp}.
Narrow emission is present in the host-galaxy component.
Since we aimed to analyse the observed narrow line emission, we did not subtract any narrow emission from 
the observed.
We masked the narrow lines in the host galaxy component before subtracting it (H$\beta$, \text{[OIII]} 
4959\,\AA, \text{[OIII]} 5007\,\AA, H$\alpha$, \text{[NII]} 6548\,\AA, \text{[NII]} 6583\,\AA, 
\text{[SII]} 6716\,\AA, and \text{[SII]} 6731\,\AA). 
We refer to the resulting spectrum hereafter as the `decomposed AGN spectrum'.

We fitted the full wavelength range of each decomposed AGN spectrum with a power law continuum, and a 
collection of broad and narrow Gaussians.
A single Gaussian was fitted to the following narrow emission lines: the \text{[OIII]} doublet; the 
\text{[NII]} doublet; and the \text{[SII]} doublet. 
The H$\beta$ and H$\alpha$ were fitted with two Gaussians representing both a broad and narrow component. 
We distinguished between broad and narrow emission lines by constraining their Gaussian line widths. 
We constrained the broad lines to have velocity width >\,1200 km\,s$^{-1}$, and narrow lines 
<\,1200 km\,s$^{-1}$ \citep{Hao05}.

It is also known that the NLR is stratified based on ionisation, with the high ionisation lines arising 
from the inner NLR, and the low ionisation lines arising from the outer NLR \citep[e.g.][]{Veilleux91}. 
As a result, we constrained the velocity and width of the narrow lines in ionisation groups. H$\beta$ and 
H$\alpha$ are one group; the \text{[NII]} doublet and the \text{[SII]} doublet are another; and the 
\text{[OIII]} doublet forms the last group. 
Narrow lines within groups were constrained to have identical width and velocities. 
The width and velocities of the broad H$\beta$ and H$\alpha$ were left as free parameters.

Because of extra light travel time from the continuum source to the NLR relative to the BLR, we did not 
expect any intrinsic variability in the narrow emission. 
Hence, the repeat decomposed AGN spectra of each object were fitted simultaneously\footnote{
  With the exception of the BOSS spectrum of J0002-0027.
}, while constraining the narrow lines to be equal across repeat spectra. 
Broad emission and the power law AGN continuum were left free to vary across repeat spectra\footnote{
  The BOSS spectrum of J0002-0027 was fitted independently from its earlier epoch spectra as it was 
  observed with a 2" fibre rather than a 3" fibre, so we expected a weaker galaxy contribution to the 
  narrow emission (as explained in Sect. \ref{sec:decomp}).
}.

\begin{figure*}
	\includegraphics[width=\textwidth]{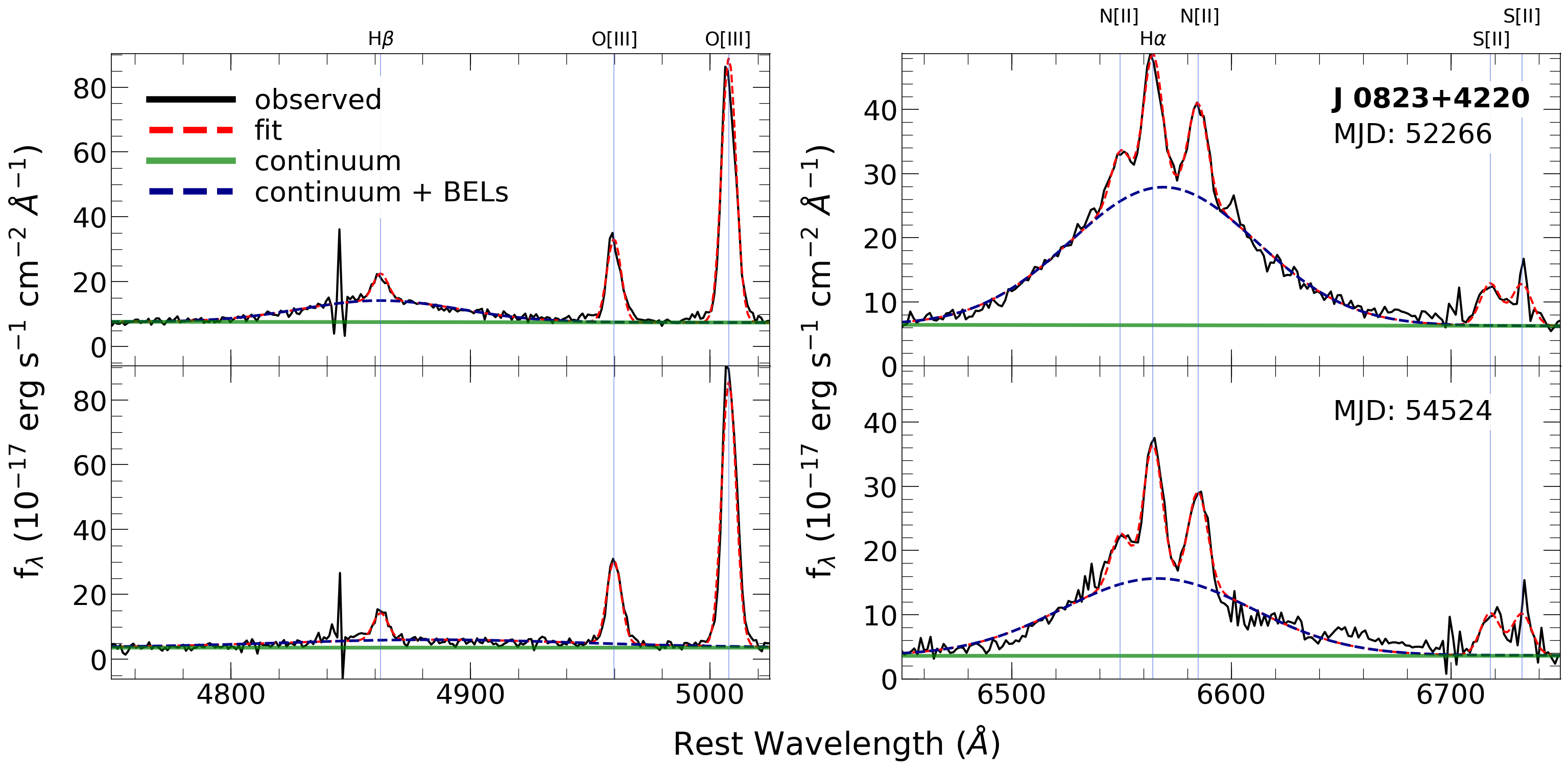}
    \caption{
      Emission line fitting to the decomposed AGN spectrum (black) of the CLQ J0823+4220. 
      The AGN spectrum was fitted (red) with a power law continuum (green) and a selection of narrow and 
      broad Gaussians. 
      The fitting procedure is described in Sect.~\ref{sec:lines}.
    }
    \label{fig:9247_lines}
\end{figure*}

The measured broad H$\beta$ and H$\alpha$ properties are listed in Table \ref{tab:CLQ_stats}. 
The best fit to the decomposed AGN spectrum (dashed red curve) of J0823+4220 is shown in Fig. 
\ref{fig:9247_lines} as an example. 
The spectra, including emission lines, were well fitted for the entire sample with two minor discrepancies. 

Firstly, the power law continuum fails to represent the full level of detail in the wavelength range 
\numrange{3400}{4500}\,\AA.
Secondly, due to the use of a single Gaussian for the broad lines, asymmetries observed in the broad lines of the 
`on-state' spectra are not fully fitted in the wings of the line (e.g. bright-state H$\beta$ and H$\alpha$ broad 
lines of J1723+5504, both spectra of J0829+4154, and in the bright-state H$\alpha$ broad line of J1358+4934). 
However, for our purpose of measuring the broad and narrow line velocity dispersion and fluxes, the fits are acceptable.

\subsection{Black hole mass and Eddington ratios}
\label{sec:bhmass}

From measurements of the decomposed AGN spectrum continuum and emission lines made in Sect. 
\ref{sec:lines}, we can estimate the black hole mass M$_{\text{BH}}$ and the Eddington 
ratio $\lambda_{edd} = L_{\text{bol}} / L_{\text{Edd}}$ of the six CLQs. 
We used H$\alpha$ for the black hole mass measurement because four of the six CLQs retain a relatively 
strong broad H$\alpha$ compared to the H$\beta$ in their dim-state spectra. 
Therefore, the H$\alpha$-derived black hole mass was better for use as a comparative measure across the 
repeat spectra. 
We used the black hole mass the equation from \citet{Greene10}:

\begin{multline}
	\text{M}_{\text{BH}} = 9.7 \times 10^6 \left[ \frac{\text{FWHM}(\text{H} \alpha)}{1000\, \text{km}\, \text{s}^{-1}} \right] ^{2.06}\\
    \times \left[ \frac{\lambda L_{5100}}{10^{44} \, \text{erg}\,\text{s}^{-1}} \right] ^{0.519} \: \text{M} _{\sun} .
	\label{eq:BHmass}
\end{multline}

Using the FWHM of the H$\alpha$ broad emission and the monochromatic rest 5100\,\AA\:AGN luminosity 
$\lambda L_{5100}$ from the fitted power law continuum.

The bolometric luminosity $L_{\text{bol}}$ of the CLQs was estimated by applying the linear bolometric 
correction of 8.1 from \citet{Runnoe12} to the measured $\lambda L_{5100}$. 
The Eddington ratio was then $L_{\text{bol}} / L_{\text{Edd}}$ where $L_{\text{Edd}}$ was found from the 
H$\alpha$-estimated BH mass from equation \ref{eq:BHmass}, as 
$L_{\text{Edd}} = 1.25 \times 10^{38} \text{M}_{\text{BH}} / \text{M} _{\sun}$. 
All inferred AGN properties are listed for each repeat observation in the sample in Table \ref{tab:CLQ_stats}.

The CLQs have black hole masses in the range \numrange{7e7}{7.2e8}\,$\text{M}_{\sun}$. 
Black hole masses derived for each object across the repeat spectra are within 0.2\,dex generally 
consistent within the uncertainties. 
The CLQs have Eddington ratios in the range \numrange {0.006} {0.04}. 
This shows that all six CLQs are hosted by massive black holes, with accretion rates well below the 
Eddington limit.

\subsection{Narrow emission line analysis}

\begin{table*}
	\centering
	\caption{Measured and inferred CLQ properties.}
	\begin{threeparttable}
	\label{tab:CLQ_stats}
  \small  
  \begin{tabularx}{\textwidth}{l *{9}{c}}
		\hline
		    Object & \multirow{2}{*}{MJD} & H$\beta$ Flux & H$\beta$ FWHM & H$\beta$ z & H$\alpha$ Flux & H$\alpha$ FWHM  & H$\alpha$ z & M$_{\text{BH}}$ & \multirow{2}{*}{$\lambda_{\text{Edd}}$}\\
        (SDSS J) &  & (erg\,s$^{-1}$) & (km\,s$^{-1}$) & (km\,s$^{-1}$) & (erg\,s$^{-1}$) & (km\,s$^{-1}$) & (km\,s$^{-1}$) & ($10^8 \, \text{M}_{\sun}$) & \\
		\hline
        0823\,$+$\,4220 & 52266 & 429$\pm$20 & 3913$\pm$169 & -44$\pm$100 & 2458$\pm$25 & 4842$\pm$62 & 218$\pm$12 & 1.22$\pm$0.3 & 0.013$\pm$0.001 \\
         & 54524 & 227$\pm$44 & 5073$\pm$2252 & 309$\pm$933 & 1198$\pm$26 & 4453$\pm$150 & 62$\pm$45 & 0.729$\pm$0.050 & 0.012$\pm$0.001 \\
        1723\,$+$\,5504 & 51813 & 350$\pm$13 & 8891$\pm$517 & 1326$\pm$221 & 1456$\pm$33 & 6275$\pm$143 & 227$\pm$65 & 3.72$\pm$0.17 & 0.013$\pm$0.001 \\
         & 51997 & 509$\pm$9 & 5317$\pm$170 & 211$\pm$61 & 2551$\pm$23 & 6404$\pm$121 & 203$\pm$27 & 4.92$\pm$0.18 & 0.016$\pm$0.001 \\
        0126\,$-$\,0839 & 52163 & 296$\pm$18 & 4200$\pm$312 & -181$\pm$107 & 980$\pm$16 & 3591$\pm$73 & -9$\pm$25 & 1.22$\pm$0.05 & 0.043$\pm$0.002 \\
         & 54465\tnote{a} & 0 & ... & ... & 0 & ... & ... & ... & ...\\
        0829\,$+$\,4154 & 52266 & 1210$\pm$201 & 8061$\pm$1001 & 1597$\pm$163 & 8241$\pm$309 & 7297$\pm$158 & 1045$\pm$15 & 3.11$\pm$0.87 & 0.006$\pm$0.001 \\
         & 54524 & 2194$\pm$205 & 8201$\pm$696 & 1512$\pm$150 & 15167$\pm$45 & 7231$\pm$25 & 846$\pm$12 & 3.92$\pm$0.58 & 0.008$\pm$0.001 \\
        0002\,$-$\,0027 & 51791 & 404$\pm$18 & 12203$\pm$277 & 2334$\pm$246 & 1646$\pm$38 & 6835$\pm$281 & -304$\pm$58 & 5.47$\pm$0.46 & 0.014$\pm$0.001 \\
         & 52559 & 362$\pm$17 & 13709$\pm$618 & 3986$\pm$86 & 1105$\pm$31 & 8628$\pm$340 & -427$\pm$93 & 7.19$\pm$0.58 & 0.007$\pm$0.001 \\
        \it{(BOSS)}  & 55477 & 1199$\pm$28 & 18931$\pm$287 & 326$\pm$135 & 1655$\pm$19 & 7723$\pm$116 & -1119$\pm$30 & ... & ... \\
        1358\,$+$\,4934 & 53438 & 18$\pm$9 & 1793$\pm$1166 & 2873$\pm$551 & 201$\pm$22 & 5753$\pm$541 & 492$\pm$279 & 0.378$\pm$0.07 & 0.002$\pm$0.001 \\
         & 54553 & 238$\pm$20 & 4527$\pm$546 & 359$\pm$131 & 870$\pm$13 & 3368$\pm$92 & 148$\pm$22 & 0.282$\pm$0.02 & 0.015$\pm$0.001 \\
    \hline
	\end{tabularx}
  \tablefoot{
    H$\beta$ (H$\alpha$) Flux, FWHM, and z are the flux, FWHM and redshift with respect 
    to the H$\beta$ (H$\alpha$) rest wavelength, respectively, of the H$\beta$ 
    (H$\alpha$) broad component, M$_{\text{BH}}$ is the black hole mass calculated 
    using the relation of \citet{Greene10} and the broad H$\alpha$ FWHM, and 
    $\lambda_{\text{Edd}}$ is the Eddington ratio. 
    All stated errors are 1$\sigma$ uncertainties generated from a 10$^2$ Monte Carlo 
    simulation of the emission line fitting procedure from Sect. \ref{sec:lines}.\\
    \tablefoottext{a}{
      The AGN component of J0126-0839 completely dimmed in its second epoch spectrum. 
    }
  }
	\end{threeparttable}
\end{table*}

\begin{figure}
	\includegraphics[width=\columnwidth]{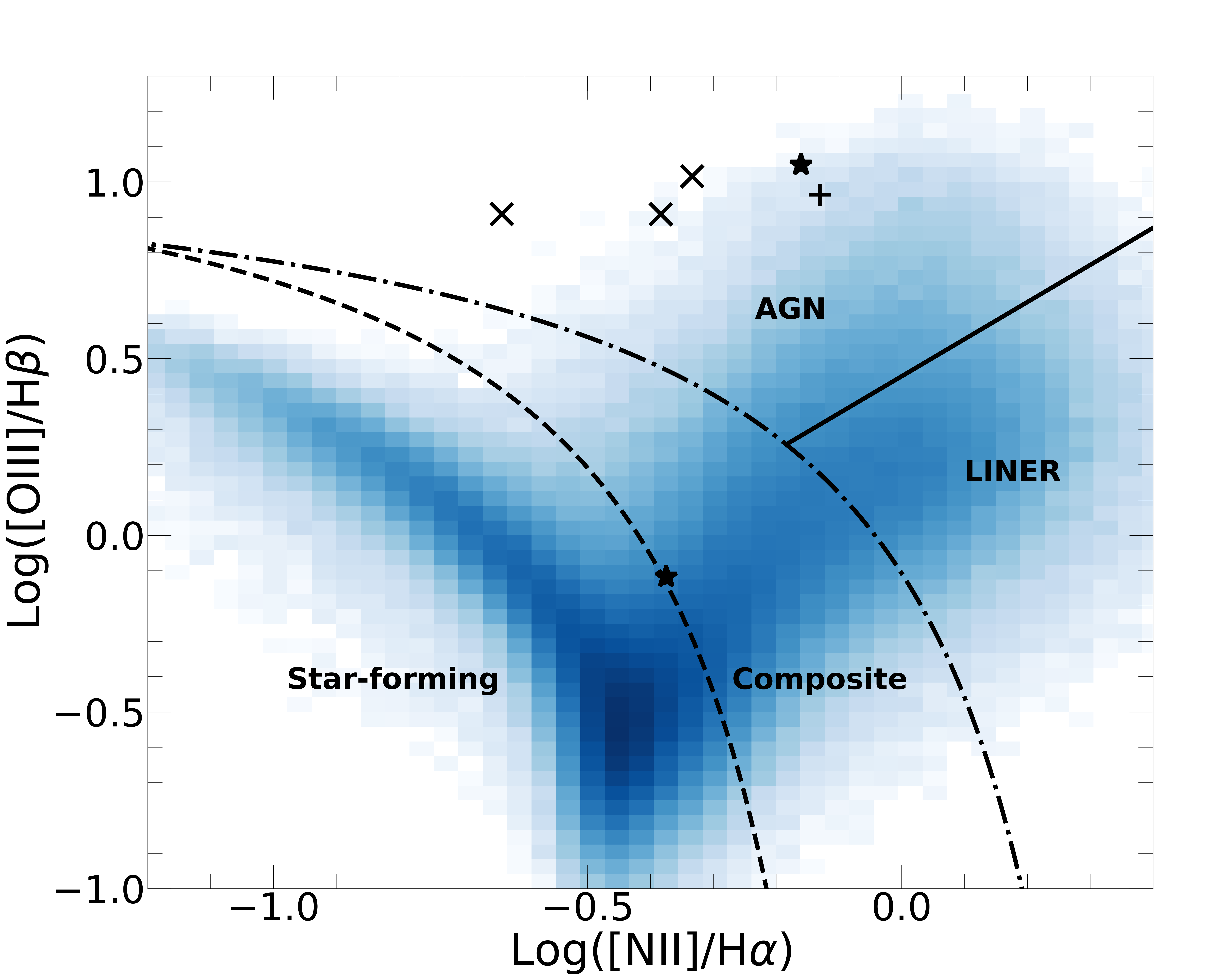}
    \caption{
      BPT diagram showing the narrow emission line ratios of the six CLQs, three turning on 
      (cross) and two turning off (star), and one turning off then on (plus). 
      The heat map was produced using the MPA-JHU emission line analysis of 927,552 DR7 objects, 
      thereby showing the overall SDSS trend \citep{SDSS_BPTdata}. 
      Three emission line classification schemes are shown. 
      \citet{Kauffmann03} (dashed line) separates star-forming and composite host galaxies, 
      \citet{Kewley06} (dotted line) separates AGN and composite host galaxies, and \citet{Schawinski07} 
      separates LINERs from AGN.
    }
    \label{fig:BPT}
\end{figure}

The size of the NLR means that light travel time from the continuum source to the NLR is 
$\sim10^{\numrange{3}{4}}$ years, compared to $\sim10^{\numrange{1}{2}}$ days to the BLR 
\citep[e.g.][]{Antonucci93, UrryPadovani1995}. 
Narrow line emission can therefore constrain the photoionisation mechanism of the NLR gas 
over the last $10^{\numrange{3}{4}}$ years. 
The observed $\text{[OIII]}$/H$\beta$ and $\text{[NII]}$/H$\alpha$ line ratios, measured as 
described in Sect. \ref{sec:lines}, of all the CLQs are shown in Fig. \ref{fig:BPT} on 
a BPT diagram \citep{BPT81}. 
We include the BPT classification schemes of \citet{Kauffmann03} and \citet{Kewley06} to 
distinguish between objects whose emission is dominated by either AGN or star formation. 
The AGNs are further categorised into Seyfert AGNs and low ionisation nuclear emission 
line regions (LINERs) by the scheme of \citet{Schawinski07}.

Fig. \ref{fig:BPT} shows that all of the CLQs' emission lines were ionised either fully 
or partially by an AGN continuum, including J0126-0839, which is classified in the composite region. 
Its emission line ratios are likely caused by strong star formation in its host galaxy. 
\citet{Kauffmann03} estimate that an AGN continuum contributes between 30 and 90 per cent of the 
ionising emission for objects in this region. 
NLR ionisation by an AGN disfavours the TDE scenario proposed by \citet{Merloni15}. 
They propose that the changing-look behaviour exhibited by the object identified by \citet{LaMassa15} 
(J0159+0033) was produced by observation of the object during and after a luminous flare caused by the 
tidal disruption of a star passing by a supermassive black hole. 
For such an observation, we expect narrow line ratios more indicative of star-formation. 
Following \citet{Ruan16}, we also examined the flux of the strongest narrow line, \text{[OIII]} 
5007\,\AA, to further test the plausibility of the TDE scenario. 
Measured \text{[OIII]} 5007\,\AA\,line fluxes of the sample yielded luminosities in the range 
\numrange{0.8}{25}\:$\times 10^{41}$erg\,s$^{-1}$. 
This is in contrast with fluxes measured in typical TDEs, which have significantly fainter 
\text{[OIII]} 5007\,\AA\, emission. 
It is unlikely, then, that the observed variability in these CLQs was caused by a TDE 
\citep[e.g.][]{Gezari06,Gezari09,Gezari12}.

The NLR in all three turning on CLQs have been ionised by an AGN-like continuum source. 
If we assume that in the `off-state' the AGN continuum emission has dimmed significantly, 
this implies that all three objects have not remained in the off-state for the last $\sim10^{\numrange{3}{4}}$ 
years, but instead have been in a state of flickering. 
The flickering could be an important part of AGN evolution. 
If we assume an equal likelihood of observing a flickering CLQ in the on or off state, that 
idea is encouraged by the identification of an equal number of turning off and turning on CLQs.


\section{Discussion}
\label{discuss}

\begin{table}
	\centering
	\caption{
    Number of CLQs that were observed turning-off, turning-on, and both.
  }
	\label{tab:on_off}
    \begin{tabularx}{\columnwidth}{ l *{3}{Y} }
		\hline
		\multirow{2}{*}{Paper} & \multicolumn{3}{c}{CLQ behaviour} \\
        & on & off & both \\
        \hline
        This paper & 3 & 2 & 1 \\
        1 & 0 & 3 & 0 \\
        2 & 4 & 5 & 1 \\
        3 & 16 & 10 & 0 \\
		\hline
	\end{tabularx}
  \tablebib{
    (1) \citet{Ruan16}; (2) \citet{MacLeod16}; (3) \citet{Yang18}.
  }
\end{table}

\begin{figure*}
  \includegraphics[width=\textwidth]{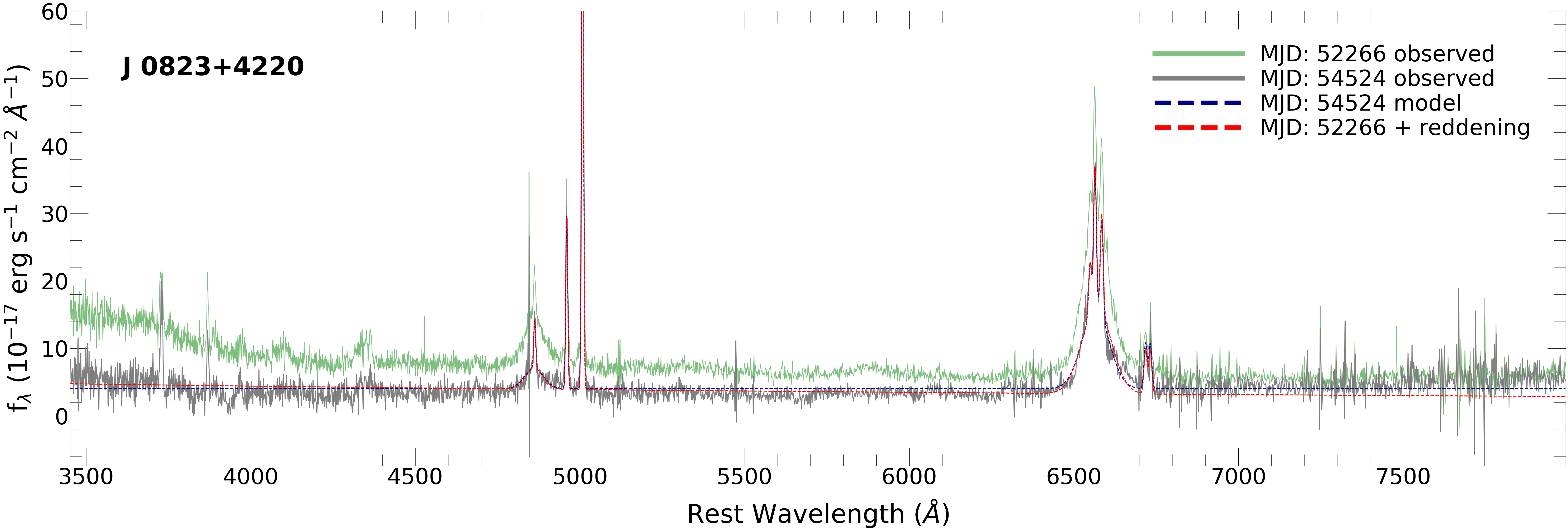}
    \caption{
      Comparison of the reddened bright-state spectrum with the directly fitted dim-state spectrum for 
      the CLQ J0823+4220. 
      The dim- (grey curve) and bright-state (green curve) observed spectra are shown. 
      We reddened the fitted components (as per Sect. \ref{sec:lines}) of the bright-state spectrum 
      to fit the dim-state spectrum. 
      This produced the red, dashed curve with a best-fit $E(B\,-\,V) = 0.176$. 
      We also show the spectral components fitted directly to the dim-state spectrum as the blue, 
      dashed curve for comparison.
    }
  \label{fig:Redden}
\end{figure*}

We have used difference spectra to search for CLQs in SDSS DR7 without pre-selecting on photometric 
variability or a change in spectral classification in the pipeline. 
We identified six CLQs, four of which are newly discovered using the difference spectrum method developed here. 
This shows that the approach presented can be very effective for finding CLQs and could be extended to higher 
redshift by using different broad lines or to different spectroscopic datasets, including SDSS-BOSS spectra.

\subsection{Comparison to other CLQ searches}

There have been three systematic searches that yielded a total of 50 CLQs to date 
\citep{Ruan16,MacLeod16,Yang18,MacLeod19}. Previous searches relied on changes in object classification or large photometric variability to select CLQs. Here, we compare the properties of the previously selected CLQs to those selected here using difference spectra.

\citet{Ruan16} first identified the CLQ J0126-0839, which is also recovered in this paper, from the object's SDSS-II spectra. 
They showed that the object, whose classification transitioned from galaxy-like to quasar-like, is very likely to display changing-look behaviour. 
\citet{Ruan16} included BOSS spectra (SDSS DR12) in their search, which yielded one additional new CLQ, J233602.98+001728.7 (hereafter referred to as J2336+0017), and the recovery of J0159+0033, which was first identified by \citet{LaMassa15}. 
The identified CLQs have luminosities $42.30 \leqslant$ log$_{10}(\lambda L_{5100}) \leqslant 43.56$ with a 
median of 43.43, comparable to the luminosities in our sample. 

\citet{MacLeod16} propose that significant broad emission line (BEL) variability will be associated with at least a 1.0 mag variability of the \textit{g}-band photometry of a CLQ. 
They used repeat photometry from SDSS and Pan-STARRS1 to select CLQ candidates, and confirmed nine new CLQs, spanning the redshift range 
0.204 $\leqslant z \leqslant$ 0.625, using repeat spectroscopy in both SDSS-II and BOSS (SDSS-III). 
\citet{MacLeod19} extended this work by expanding their photometric search to include highly variable objects in SDSS-II that did not possess any BOSS spectra. 
They confirmed 17 new CLQs using optical spectroscopy from the William Herschel, MMT, Magellan, and Palomar telescopes. 
The CLQs in this sample have luminosities spanning the luminosity range $44.15 \leqslant$ log$_{10}(\lambda L_{5100}) \leqslant 44.45$ for a median value 
of 44.34, approximately an order of magnitude higher than those found in this study. 
\citet{MacLeod19} note that their CLQs are at lower Eddington ratio relative to the overall quasar population in SDSS.

Since we do not use BOSS in our sample selection, we do not recover the CLQs identified by \citet{MacLeod16}. 
Upon passing the additional BOSS spectra through our sample selection algorithm in a separate test, 
seven of the total ten CLQs listed by \citet{MacLeod16} are recovered. 
The un-recovered objects either have redshift $z > 0.6$, do not possess a visible H$\alpha$ BEL, or have missing data in
one of its spectra.

\citet{Yang18} conducted three different systematic searches for CLQs. 
The first directly measured cross-matched LAMOST and SDSS spectra with fitted Balmer BELs. 
They identified ten new CLQs from this search. 
Their second search was of SDSS DR14 spectra relying on SDSS pipeline classification in the same way as \citet{Ruan16}. 
They identified nine CLQs, including five new CLQs and four previously identified sources. 
Of the five new CLQs found by the pipeline classification search, three are identified by \citet{Yang18} only from repeat SDSS Legacy spectra, as in this paper. 
All three of these CLQs are recovered by our search method, but we list only J1358+4934 for displaying clear and strong changing-look behaviour. 
The luminosities of the CLQs in this sample span the range $42 \leqslant$ log$_{10}(\lambda L_{5100}) \leqslant 44.48$. 
Their final search leveraged photometry similarly to \citet{MacLeod16} and \citet{MacLeod19}.

One point of comparison for each CLQ sample is the ratio of turning-on to turning-off objects. 
In this paper, as in previous work, approximately an equal fraction of objects turn on and off. 
Combined with the prevalence of AGN ionisation in the narrow line region, this supports flickering CLQs, as opposed to long time scale transitions between states.

Our sample covers a wide range of luminosities, with 
$43.24 \leqslant$ log$_{10}(\lambda L_{5100}) \leqslant 44.08$, as listed in Table \ref{tab:changinglook}, and a median value of 43.80. 
This search, therefore, reaches luminosities about an order of magnitude lower than photometric searches \citep{MacLeod16}. 
Additionally, the method used here recovers objects not found when relying on a change in pipeline classification \citep{Ruan16,Yang18}. 
Hence, we have shown that the method presented here allows recovering a more complete sample of CLQs than previous methods down to lower luminosities.

\begin{table}
	\centering
	\caption{
    Fitted $E(B\,-\,V)$ values for the Calzetti extinction curve applied to the bright-state spectra 
    of the six CLQs.
  }
	\label{tab:ebv}
  \begin{tabularx}{\columnwidth}{ l YY }
		\hline
		  Object & $\Delta t_{obs}/t_{rest}$ & $E(B\,-\,V)$ \\
    \hline
      0823\,$+$\,4220 & 2258/1960 & 0.176 $\pm$ 0.002\\
      1723\,$+$\,5504 & 184/142 & 0.118 $\pm$ 0.001 \\
      0126\,$-$\,0839 & 2302/1922 & 1.833 $\pm$ 0.373 \\
      0829\,$+$\,4154 & 2258/2005 & 0.204 $\pm$ 0.001 \\
      0002\,$-$\,0027 & 758/587 & 0.157 $\pm$ 0.002 \\
      1358\,$+$\,4934 & 1115/999 & 0.526 $\pm$ 0.013 \\
		\hline
	\end{tabularx}
  \tablefoot{
    $\Delta t_{obs}/t_{rest}$ is the observed and rest frame time between spectra in days. 
    $E(B\,-\,V)$ is the extinction value required to dim the bright-state spectrum to approximately match the dim-state.
  }
\end{table}

\subsection{Possible causes of changing-look behaviour}

In this section, we consider two possible causes for changing-look behaviour in our sample: 
changes in accretion rate and variable obscuration.

\subsubsection{Variable accretion rate}

The most common explanation for changing-look behaviour is that variability in the accretion rate causes 
changes in the continuum emission as well as broad line emission. 
This  model is favoured for many CLQs \citep{MacLeod16, MacLeod19, Ruan16, LaMassa15}.
The complete disappearance of the broad line emission is predicted by \citet{Elitzur2009} at luminosities 
below $5\times 10^{39} (M_{BH}/(10^7 M_{\odot})^{2/3})$ [erg/s], consistent with the variability changes 
in a majority of CLQs \citep{MacLeod16, MacLeod19, Ruan16, LaMassa15}. 
In previous studies, variable accretion has been the favoured explanation for CLQ behaviour.

In our sample, the difference spectra of all six CLQs, shown in Figs. \ref{fig:Off}, \ref{fig:On}, and \ref{fig:Off_and_on}, 
are well fitted by disc emission slightly shallower than the standard $f_\upsilon = \upsilon ^{1/3}$ thin accretion disc model 
\citep{Shakura73}.
This result is consistent with previous analysis of difference spectra in CLQs \citep[e.g.][]{Ruan14}. 
To quantify how much shallower the continuum is compared with a standard thin disc continuum, we reddened the standard 
thin accretion disc power-law using the Calzetti extinction curve \citep{Calzetti00} with $E(B\,-\,V)$ as a free parameter. 
All our objects yielded $E(B\,-\,V) < 0.005$. 
This is below the cut-off for dust-reddened quasars of $E(B\,-\,V) = 0.04$ defined by \citep{Richards03}, 
which does not indicate significant amounts of dust reddening.
Spectral energy distributions shallower than the standard thin accretion model are expected for models including non-black body radiation 
due to scattering in the disc \citep{Hall18} as well as for models including inhomogeneities in the accretion flow \citep{Dexter11}.

\citet{Elitzur2009} propose that the broad line region that is formed in a disc wind disappears below a 
certain (black hole mass dependent) accretion rate. 
\citet{MacLeod19} show that their CLQs could be explained by the \citet{Elitzur2009} model. 
However, we found our sources to lie at least 1 order of magnitude above the cut-off. 
While accretion disc variability matches our data, they are not consistent with the complete disappearance 
of the disc seen by \citep{MacLeod19} and predicted by \citet{Elitzur2009}. 
In this model, our objects are explained by strong changes in the accretion rate, however, we do 
not expect a full disappearance of the broad line region, as evident by the remaining H$\alpha$ emission 
in dim-state spectra.

\citet{NodaDone18} propose that the accretion disc spectrum shows a transition between a hard and soft 
state at $L/L_{edd}\sim0.02$, the change in the accretion disc spectrum occurs primarily in the X-rays, 
changing the number of ionising photons and therefore the strength of the broad line region. 
Their transition luminosity is broadly consistent with the Eddington ratios measured here and can, 
therefore, explain changing-look behaviour in this sample.

Higher time-sampling would be needed to model the changing-look behaviour in our sample in more detail, but the difference spectra as well as 
the Eddington ratio regime found are consistent with a change in accretion rate leading to a dimming or disappearance of broad lines. 
Variable accretion can therefore explain our CLQs.

\subsubsection{Variable obscuration}
\label{sec:obscure}

Variable obscuration could cause changing-look behaviour if an isolated dust cloud outside of the 
broad line region obscures both the continuum and broad line emission. 
This model has been disfavoured in previous studies due to the short timescales of CLQ behaviour \citep[e.g.][]{LaMassa15}. 
Nevertheless we tested if variable obscuration can explain changing-look behaviour in our sample. 

To test this scenario, we fitted the dim-state spectrum with a reddened version of the bright-state AGN spectrum (power-law and broad lines), 
keeping the host galaxy fixed and leaving only the reddening $E(B\,-\,V)$ as a free parameter (see Sect. \ref{sec:decomp} for details on the decomposition). 
We used the Calzetti extinction curve with $R_V = 4.05$. The reddening fit is performed for all six CLQs in the sample, including J0126-0839 which has no 
AGN component in the dim-state spectrum. 
The best fit $E(B\,-\,V)$ ranges from $\sim$\;0.1-1.8 for the sample (see Table \ref{tab:ebv}). 
Fig. \ref{fig:Redden} shows the fit for J0823+4220. 
Both the continuum and broad line strength are reproduced by the reddening fit over the full wavelength range. 
The reddened model (dashed red line) almost exactly matches the continuum variability model in 
Sect. \ref{sec:lines}. 
The continuum and broad line emission variability is well modelled by obscuration in all but one object. 
J0829+4154 shows excess emission at $\lambda \sim$ \numrange{3500}{3700}\,\AA\, that is not well fitted 
by the reddening model. 
The broad emission on the other hand is well reproduced, with both the broadening and dimming from 
bright-state to dim-state accounted for. 

The spectral changes in five of the six objects in our sample are fitted well by changes in obscuration, in conflict with other CLQ searches \citep{Ruan16, Runnoe16,MacLeod16, MacLeod19}. 
This might indicate an inherent difference in the lower luminosity sources found using our search method.

While the spectral changes are well fitted by variable obscuration, the timescales of the observed also need to be explained. 
\citet{LaMassa15} analysed dust obscuration as a source of CLQ behaviour and reject this explanation due to the time scales of the variations. 
The crossing timescale for a dust cloud, $t_{\text{cross}}$, is given by:

\begin{equation}
  t_{\text{cross}} = 0.07 \left( \frac{r_{\text{orb}}}{1 \text{lt-day}} \right)^{3/2} \left( \frac{M_{\text{BH}}}{10^8 M_{\odot}}\right)^{-1/2} \text{arcsin} \frac{r_{\text{src}}}{r_{\text{orb}}}\; \text{years}
\end{equation}

with $r_{\text{orb}}$ the orbital radius in light-days, $r_{\text{src}}$ the orbital radius of the source 
and $M_{\text{BH}}$ the black hole mass. 
For sources studied by \citet{LaMassa15}, this is $\sim$\;10-20 years, an order of magnitude higher than the observed variability timescales. 
\citet{Runnoe16} also found similar discrepancies in the timescales. 

Using black holes masses from Sect. \ref{sec:bhmass} and assuming an obscurer just outside the broad line region with a radius calculated 
from the luminosity using \citep{Bentz2009}, we get a crossing time as low as $\sim$\;3 years, consistent with our objects. 
However, for the orbital radii consistent with the same timescale, any obscurer would have to have a size comparable to the broad line region 
to explain the observed behaviour, if the obscurer is located at a larger radius, the timescale increases. 

Therefore we conclude that for our objects, while obscuration reproduces the changes in spectral shape observed, the timescales of the observed 
CLQ behaviour remain difficult to explain. Changes in accretion rate are more likely to be the cause for the changing-look behaviour.


\section{Conclusions}
\label{conc}

In this paper, we used repeat spectra of 24,782 extra-galactic objects in SDSS-II \citep{Abazajian09} 
to search for CLQs with variability levels weaker than those detected in previous studies. 
We selected all sources that show weakening or strengthening of continuum emission bluewards of 4000\;\AA\:as 
well as disappearance or appearance of the H$\beta$ or H$\alpha$ broad emission lines between SDSS epochs. 
As a result, we detected six CLQs, four of which are newly discovered here.

By detecting new CLQs in a previously well-studied spectrometric dataset, we have shown that using direct 
comparison between repeat spectra is effective for detecting CLQs. Furthermore, the method can be used to 
identify CLQs that lie below the detection threshold of methods that rely on pre-selection either by 
photometric variability or change of spectral type in pipeline classifications.

Our other findings can be summarised as follows. 
Firstly, the six CLQs have redshifts 0.1 $\leqslant$ z $\leqslant 0.3$. 
Two show transitions with disappearing, three with appearing, and one both disappearing and re-appearing 
broad emission lines. 
The CLQ J1723+5504 transitioned from an intermediate type 1.8 AGN to a type 1 AGN in 184 days (142 in the rest frame); 
the fastest observed changing look behaviour in a distant AGN with quasar luminosity.

Secondly, we decomposed the CLQ spectra by fitting galaxy templates, a power law continuum as well as broad 
and narrow line emission. 
The difference spectrum of all six CLQs was well fitted by disc emission slightly shallower than the standard 
$f_\upsilon \propto \upsilon^{1/3}$ accretion disc model, in agreement with previous studies \citep{Ruan14}.

Thirdly, the CLQs have black hole masses in the range \numrange{3e7}{4e8} $\text{M}_{\odot}$ and Eddington 
ratios $L_{\text{bol}}/L_{\text{Edd}} \sim 0.002-0.03$, making these moderately luminous AGN 
($L_{\text{bol}} \sim 10^{43-44}$ erg/s). 
Therefore, the CLQs are hosted by massive black holes with accretion well below the Eddington limit.

Fourthly, the narrow line ratios for five of the six CLQs are indicative of ionisation by an AGN. 
Only one object shows a mixture of star formation and AGN activity as the source of ionisation. 
This indicates that the ionising radiation in CLQs was dominated by AGN over the last $>10^{3-4}$ years. 
This supports that even turning on CLQs have experienced flickering in the past, rather than turning on over 
the past years.

Finally, we showed that the changes in spectral shape are explained by changes in accretion rate as well as 
changes in obscuration. 
An obscuration origin is unlikely due to the short timescales on which transitions are observed. 
Therefore, changes in accretion rate are the likely the cause of the changing look behaviour in our sample. 
A complete disappearance of the broad line region as predicted \citep{Elitzur2009} by models is not favoured, 
however, our data is consistent with a switch in the accretion state, as predicted by \citet{NodaDone18}.

We have shown that using difference spectra can be used to detect CLQs not detected by photometric 
variability and therefore represents an additional method for CLQ searches. 
Using the method presented here, we detected CLQs an order of magnitude fainter than those detected 
in photometric searches and we reached higher completeness than spectroscopic searches relying on a 
change of object type in the pipeline classification.

More data will be needed to identify the physical processes that drive changing look behaviour at the 
low luminosity end. 
Further research is also needed to determine if all CLQs are driven by the same physical process.

\begin{acknowledgements}
  We thank the anonymous referee for helpful comments. We acknowledge Benjamin Pickford for his contribution to developing the sample selection method and 
  confirming the six CLQs in the sample. 

  Funding for the SDSS and SDSS-II has been provided by the Alfred P. Sloan Foundation, the Participating Institutions, the National Science Foundation, the U.S. Department of Energy, the National Aeronautics and Space Administration, the Japanese Monbukagakusho, the Max Planck Society, and the Higher Education Funding Council for England. 
  The SDSS Web Site is http://www.sdss.org/.

  The SDSS is managed by the Astrophysical Research Consortium for the Participating Institutions. The Participating Institutions are the American Museum of Natural History, Astrophysical Institute Potsdam, University of Basel, University of Cambridge, Case Western Reserve University, University of Chicago, Drexel University, Fermilab, the Institute for Advanced Study, the Japan Participation Group, Johns Hopkins University, the Joint Institute for Nuclear Astrophysics, the Kavli Institute for Particle Astrophysics and Cosmology, the Korean Scientist Group, the Chinese Academy of Sciences (LAMOST), Los Alamos National Laboratory, the Max-Planck-Institute for Astronomy (MPIA), the Max-Planck-Institute for Astrophysics (MPA), New Mexico State University, Ohio State University, University of Pittsburgh, University of Portsmouth, Princeton University, the United States Naval Observatory, and the University of Washington.
\end{acknowledgements}


\bibliographystyle{aa}
\bibliography{citations}

\end{document}